\documentclass[fleqn,usenatbib]{mnras}

\usepackage{epsfig}
\usepackage{times}
\usepackage{amsmath}
\usepackage[british]{babel}             
\usepackage[british]{babel}             
\usepackage{newtxtext}                  
\usepackage[slantedGreek]{newtxmath}    
\usepackage[T1]{fontenc}                
\usepackage{graphicx}                   

\newif\ifAMStwofonts
\newcommand{\target}{PSR\,J1023+0038}

\newcommand{\erg}{\,erg\,s$^{-1}$}

\title[Hot clumpy accretion flow in  PSR\,J1023+0038]
{Evidence for hot clumpy accretion flow in the transitional millisecond pulsar 
PSR\,J1023+0038}

\author[T.\,Shahbaz et al. ]
       {T.\,Shahbaz,$^{1,2}$\thanks{E-mail: tsh@iac.es}
        Y.\,Dallilar,$^3$         
        A.\,Garner,$^3$ 
        S.\,Eikenberry,$^{3}$ 
        A.\,Veledina$^{4,5}$ and 
        P.\,Gandhi$^{6}$ \\
$^1$Instituto de Astrof\'\i{}sica de Canarias (IAC), E-38200 La Laguna, 
Tenerife, Spain \\
$^2$Departamento de  Astrof\'\i{}sica, Universidad de La Laguna (ULL), 
E-38206 La Laguna, Tenerife, Spain \\
$^3$Department of Astronomy, University of Florida, 211 Bryant
Space Science Center, Gainesville, FL 32611, USA \\
$^4$Nordita, KTH Royal Institute of Technology and Stockholm University, 
Roslagstullsbacken 23, SE-10691 Stockholm, Sweden\\
$^5$Tuorla Observatory, University of Turku, V\"ais\"al\"antie 20, FI-21500 
Piikki\"o, Finland\\
$^6$Department of Physics and Astronomy, University of Southampton, 
Highfield, Southampton SO17 1BJ, UK \\}

\pagerange{\pageref{firstpage}--\pageref{lastpage}}
\pubyear{2014}

\begin{document} 

\maketitle 

\begin{abstract} 
\noindent 
We present simultaneous optical and near-infrared (IR) photometry 
of the millisecond pulsar PSR\,J1023+0038 during its low-mass 
X-ray binary phase. The $r'$- and $K_s$-band light curves show 
rectangular, flat-bottomed dips, similar to the X-ray 
mode-switching (active--passive state transitions) behaviour 
observed previously. The cross-correlation function (CCF) of the 
optical and near-IR data reveals a strong, broad negative 
anti-correlation at negative lags, a broad positive correlation 
at positive lags, with a strong, positive narrow correlation 
superimposed. The shape of the CCF resembles the CCF of black 
hole X-ray binaries but the time-scales are different. The 
features can be explained by reprocessing and a hot accretion 
flow close to the neutron star's magnetospheric radius. 
The optical emission is dominated by the reprocessed 
component, whereas the near-IR emission contains the emission from plasmoids in the hot 
accretion flow and a reprocessed component. 
The rapid active--passive state transition occurs 
when the hot accretion flow material is channelled onto the neutron star and is expelled from 
its magnetosphere. 
During the transition the optical reprocessing component decreases  
resulting in the removal of a blue spectral component. 
The accretion of clumpy material through the magnetic barrier of the neutron star 
produces the observed near-IR/optical CCF and variability.
The dip at negative lags corresponds to the suppression of the near-IR 
synchrotron component in the hot flow, whereas the broad positive correlation at positive lags 
is driven by the increased synchrotron emission of the outflowing plasmoids. The narrow peak 
in the CCF is due to the delayed reprocessed component, enhanced by the increased X-ray 
emission. 

\end{abstract}


\begin{figure*}
\centering
\includegraphics[width=1.0\linewidth]{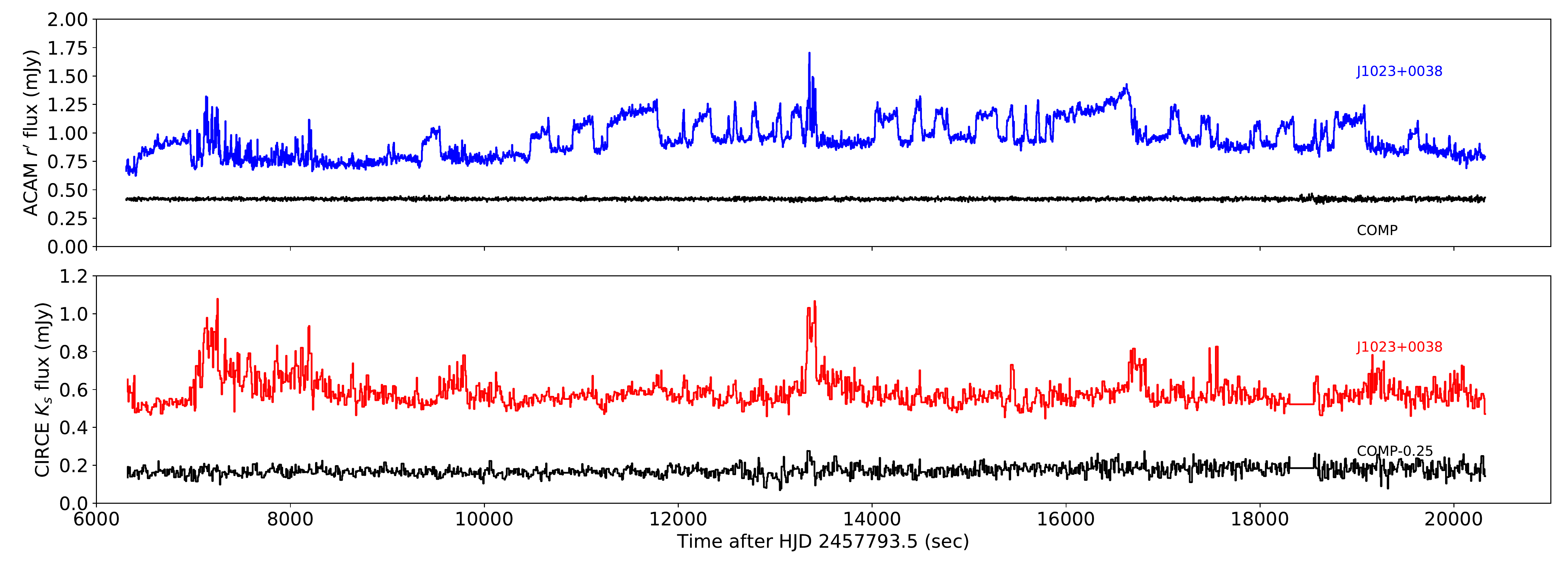}
\caption{
Observed light curves of \target\ and comparison star taken with 
the WHT+ACAM ($r'$-band) and GTC+CIRCE ($K_s$-band). The light curve of the 
($K_s$-band) comparison star has been shifted by 0.5 for clarity. 
}
\label{fig:lcurve}
\end{figure*}

\begin{figure*}
\centering
\includegraphics[width=1.0\linewidth]{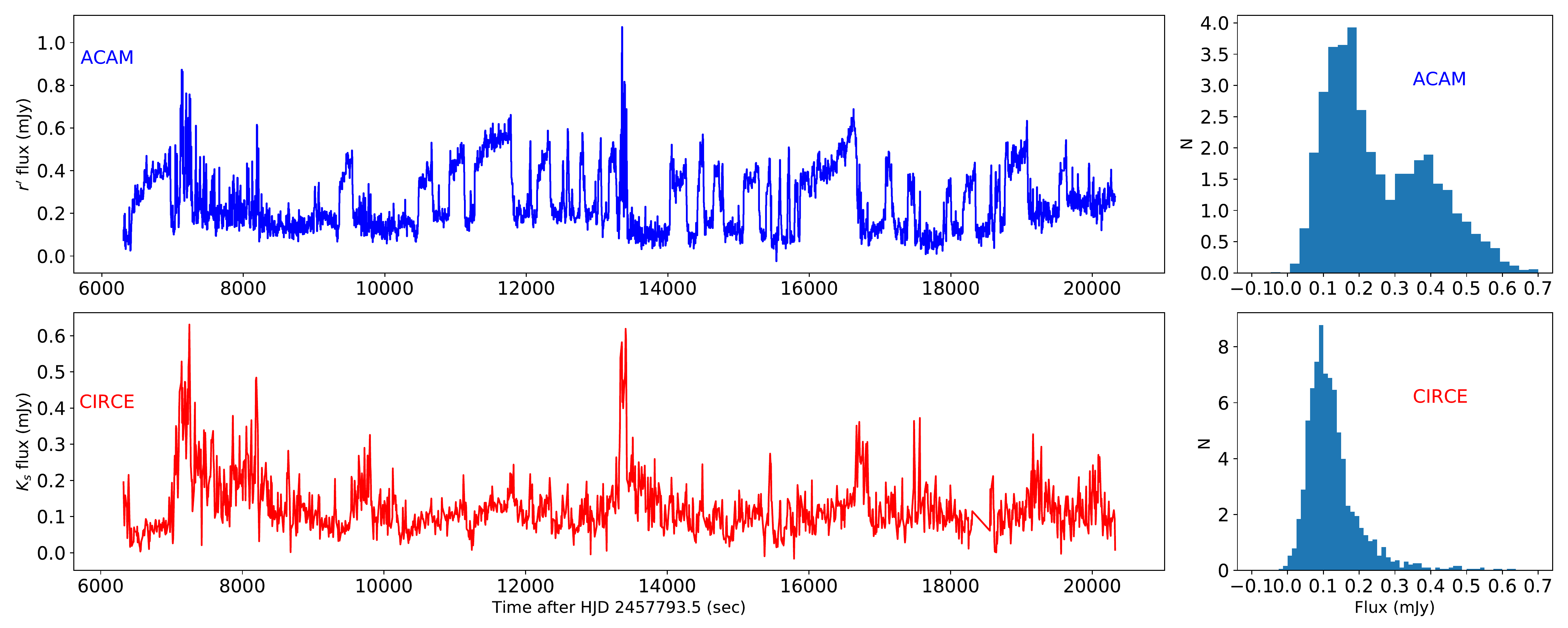}
\caption{
The de-reddened and de-trended $r'$- and $K_s$-band light curves 
of \target. The right panel in each plot shows the histogram of the flux 
values. The active- and passive-state bi-modal behaviour is clearly evident 
in the $r'$-band light curve. 
}
\label{fig:detrend}
\end{figure*}

\section{Introduction}
 
Millisecond radio pulsars (MSPs) are believed to be the descendants of 
old, slowly rotating weakly magnetic field neutron stars that show radio, 
X-ray, and/or $\gamma$-ray pulsations, that are present in low-mass X-ray 
binaries (LMXBs). The theory of recycled pulsars suggests that after the 
neutron star turns off its radio pulsations, the transfer of angular 
momentum from the low-mass companion star to the neutron star via 
accretion is responsible for the spin-up of the initially slowly rotating 
neutron star \citep{Alpar82, Radhakrishnan82}. During this accretion 
phase, X-ray pulsations at the spin frequency of the neutron star are 
detected and the system shows the typical features of an accreting 
millisecond X-ray pulsar (AMXP). The recycling scenario of MSPs discussed 
above was initially confirmed by \target\ \citep{Archibald09}, which was 
the first source that showed the transition between the rotation powered 
state to the accretion powered state. We now know of three transitional 
MSPs: \target\ \citep{Archibald09}, XSS\,J12270--4859 \citep{Bogdanov14} 
and M28I (=IGR\,J18245--2452) \citep{Papitto13}.

\target\ transitioned from a rotation-powered to an accretion-powered LMXB 
state in 2013 June \citep{Stappers13, Patruno14, Takata14} and so far has 
remained in this state. The state transition involved the disappearance of 
the radio pulsed signal as well as an increase in the GeV flux by a factor 
$\sim$\,5 \citep{Stappers14}. In the optical, the system brightened by 
$\sim$\,1\,mag and showed the presence of several broad emission lines 
typical of an accretion disc \citep{Halpern13}. In the LMXB state, the 
X-ray luminosity varied between $\sim 10^{32.5}-10^{34.5}$\erg\ showing 
three distinct states: (1) high ``active'' state with $L_X \sim 
3\times10^{33}$\erg\ present 80 per cent of the time; (2) low ``passive'' 
state with $L_X\sim 5 \times$10$^{32}$\erg\ present 20 per cent of the 
time; (3) and the ``flare'' state with $L_X \sim 3\times 10^{34}$\erg\ 
present for about 2 per cent of the time \citep{Archibald15, Bogdanov15b}. 
Coherent X-ray pulsations have been observed but appear only in the high 
state, and have been interpreted as being due to the inflowing accretion 
material heating the magnetic polar caps of rotating neutron star 
\citep{Archibald15}. The transitions between the low and high states are 
very rapid with a time-scale of 10\,s and results in rectangular-shaped 
dips in the X-ray light which seem to be ubiquitous in all of the 
X-ray observations. Similar variability has also been observed at 
optical wavelengths with rectangular dip ingress and egress times of 
$\sim$20\,s \citep{Shahbaz15, Bogdanov15a}.
However, given the limited simultaneous optical/X-ray 
observations, it is not clear if the X-ray and optical dips are 
direct counterparts of each other. Simultaneous {\it XMM--Newton} 
X-ray and $B$-band optical light curves show X-ray dips that do not 
usually have corresponding optical dips \citep{Bogdanov15b}. However, 
the uncertaintes in the $B$-band data are large and given that the 
dips in the blue are weak \citep{Shahbaz15}. one cannot rule out if 
the dips are present. 

In this paper we present the results of simultaneous optical/near-IR fast 
imaging of \target. We deterimine the auto-correlation function (ACF) of 
the optical and near-IR light curves as well as the cross-correlation 
(CCF) function of the two light curves. We explain the observed features 
of the ACFs and CCFs as due to the hot, clumpy inner accetion flow close to the 
pulsar's magnetosphere.

\begin{figure*}
\centering
\includegraphics[width=1.0\linewidth]{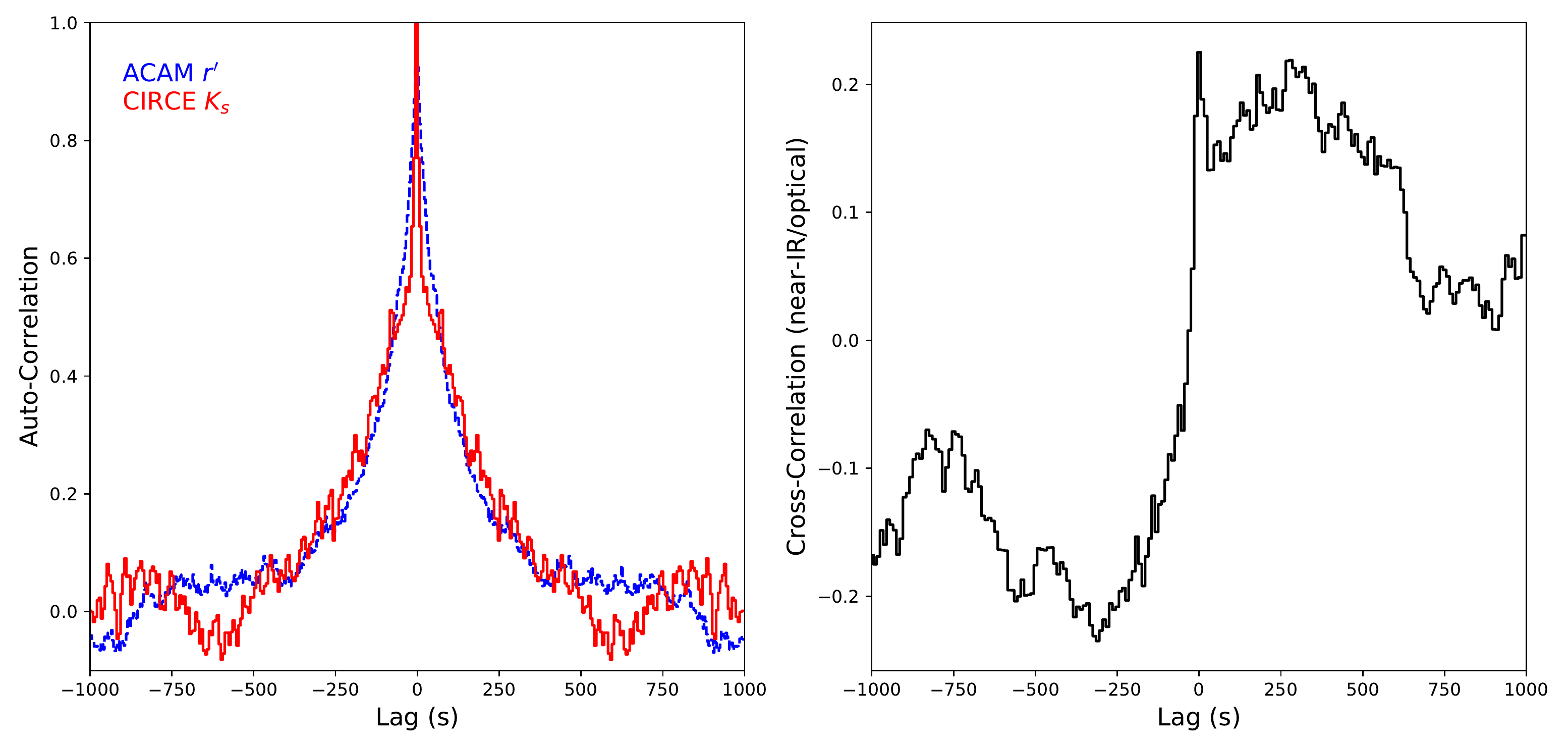}
\caption{ 
Left: ACF of the CIRCE $K_s$ /red solid line) and ACAM $r'$ (blue dashed 
line) \target\ light curves as a function of time delay. The near-IR is 
broader than the optical ACF. Right: CCF between the simultaneous $K_s$ 
and $r'$ light curves as a function of time delay. A strong 
anti-correlation is observed where near-IR light curve is delayed with 
respect to optical light curve by $\sim$300\,s.
}
\label{fig:ccfs_target}
\end{figure*}

\section{Observations and data reduction}
\label{obs}

\subsection{High-speed optical photometry}
\label{obs:optical}

Fast optical imaging of \target\ was carried out on the night of 2017 
February 8 using ACAM on the 4.2\,m William Herschel Telescope (WHT). The 
CCD was binned by a factor of 2 and windowed down to 200$\times$400 
pixels, which reduced the dead time between exposures to only 1.85\,s.  
The plate scale was 0.25 arcsec pixel$^{-1}$. We used the $r'$ filter and an 
exposure time of 2.0\,s which resulted in a time resolution of 3.85\,s.  
Observations were obtained from 2017 February 9 01:37:21 to 05:31:29 UT 
and the seeing was typically between 1.0--2.0\,arcsec.

The images were first bias-subtracted and flat-fielded.  The target counts 
were then extracted using optimal photometry with a seeing-dependent 
variable circular aperture tracking the centroid of the source on each 
image. The sky background was measured from the clipped mean in an annular 
aperture around the target.  Relative photometry of \target\ was then 
carried out with respect to the local standard star 
SDSS\,J102343.30+003819.1 $r'$\,=\,14.778. The mean magnitude of \target\ 
was $r'$\,=\,16.48.

\subsection{High-speed near-Infrared photometry}
\label{obs:ir}

Fast near-Infrared imaging of \target\ was carried out on the night of 
2017 February 8 using Canarias InfraRed Camera Experiment (CIRCE; 
\citealt{Eikenberry17}) on the 10.4\,m Gran Telescopio Canarias (GTC). The 
plate scale of the array was 0.1 arcsec pixel$^{-1}$ and the image size was 
1855$\times$1365 pixels. The $K_s$ filter was used with an exposure time of 
5.0\,s and a dead-time of 1\,s, resulting in a time resolution of 5\,s. A 
five-point dither pattern was used with 5 exposures at each 
dither position. Observations were obtained from 2017 February 9 01:12:21 
to 05:31:04 UT. Dark images were also taken with the same detector 
configuration and exposure time.

The CIRCE data reduction was carried out using the 
\textsc{SuperFATBOY}\footnote{\textsc{SuperFATBOY} is an XML-controlled 
GPU-enabled Python pipeline developed at the University of Florida} data 
reduction pipeline. We first performed the linearity correction and dark 
subtraction on all the images. A flat-field was then generated and 
applied to the science images. The sky background was determined using the 
images in each dither pattern. Finally, bad pixels and cosmic ray events 
were interpolated and the images binned by 2 pixels to improve 
signal-to-noise.

Before extractng the counts from the stars in the field the images were 
first shifted and aligned. We used optimal photometry with a 
seeing-dependent variable circular aperture tracking the centroid of the 
source on each image to determine the star counts.  The sky background was 
measured from the clipped mean in an annular aperture around the target.  
Relative photometry of \target\ was then carried out with respect to the 
local standard star 2MASSXJ10235051+0038163 with $K_s$\,=\,14.894 (the 
ACAM local standard could not be used because it occasionally fell into 
the cosmetic gap of the IR array). The mean magnitude of \target\ was 
$K_s$\,=\,15.14.

\begin{figure}
\centering
\includegraphics[width=1.0\linewidth]{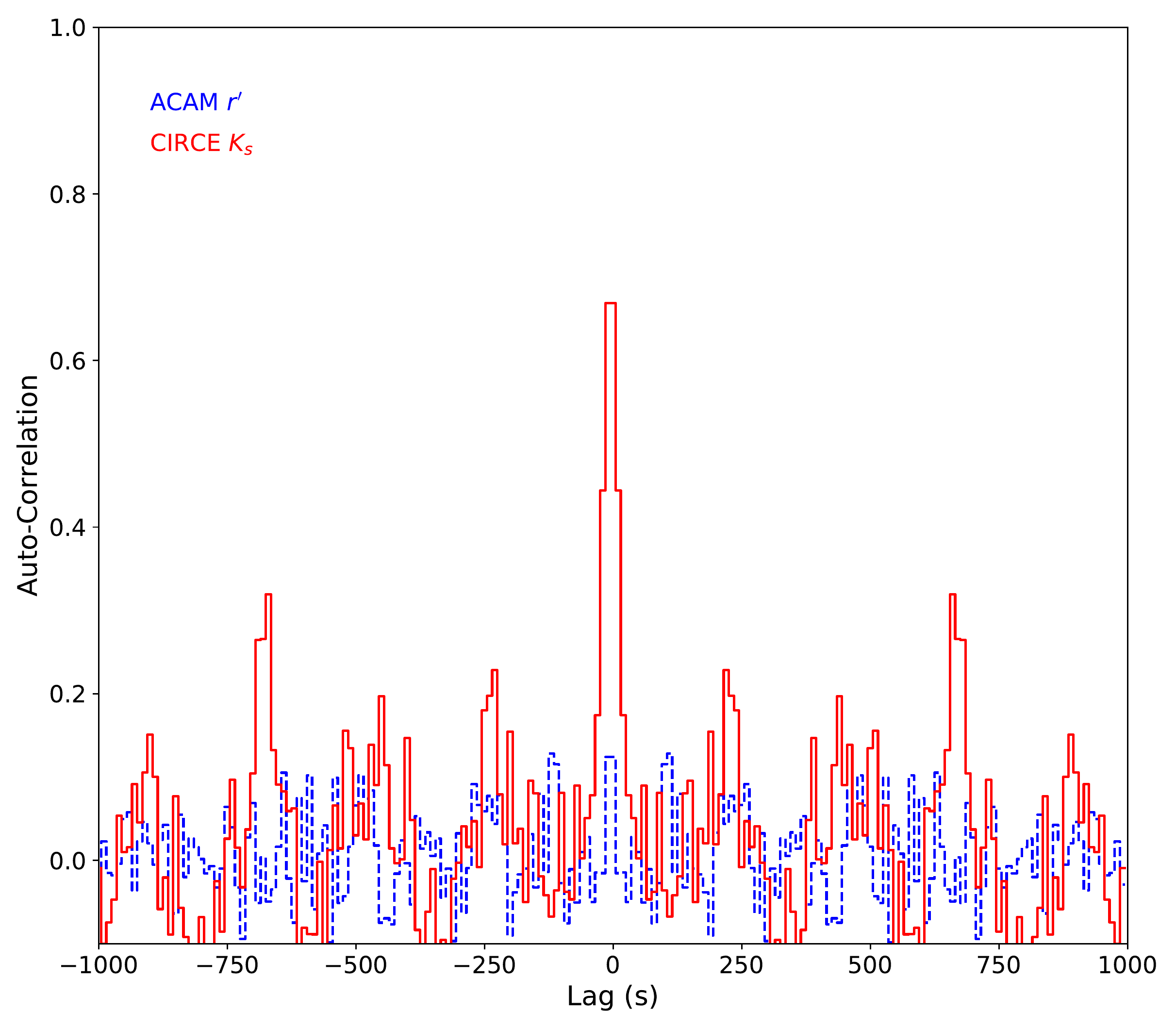}
\caption{
The ACF of the CIRCE $K_s $ (red solid line) and ACAM 
$r'$ (blue dashed line) comparison star light curves. The near-IR ACF has 
a width of $\sim$40\,s.
}
\label{fig:ccfs_cstar}
\end{figure}

\begin{figure*}
\begin{center}
  \includegraphics[width=1.0\linewidth]{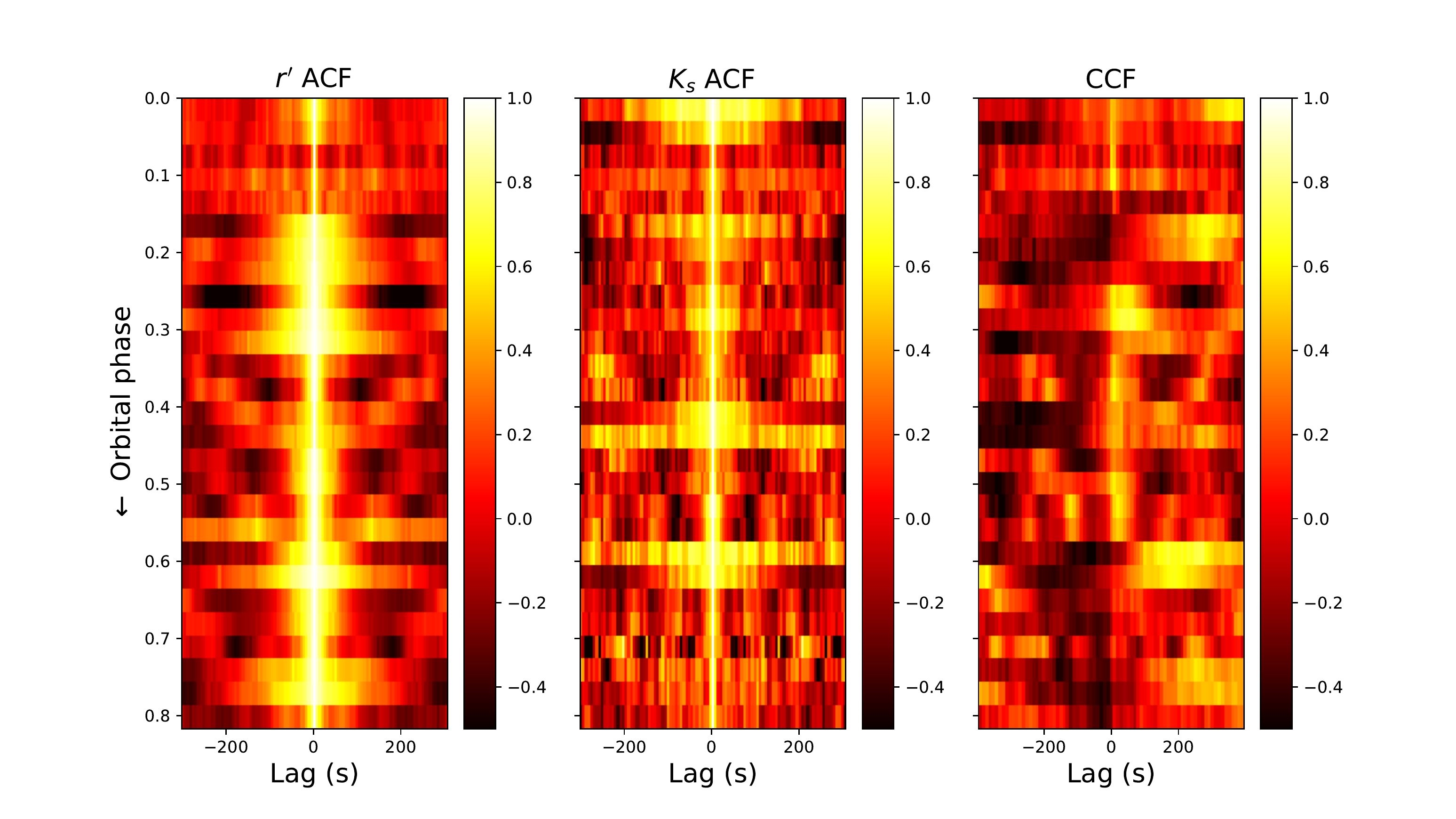}
\end{center}
\caption{
The dynamical ACF of the $r'$ (left) and $K_s$-band (middle) light 
curves and the CCFs (right) of the two light curves. We use a windows size 
of 1000\,s and a step size of 500\,s, Each horizontal line is a single 
ACF or CCF from a 500\,s segment light curve. The orbital phase ephemeris 
is taken from \citet{Shahbaz15}. 
}
\label{fig:dyn_acf}
\end{figure*}

\begin{figure*}
\centering
\includegraphics[width=1.0\linewidth]{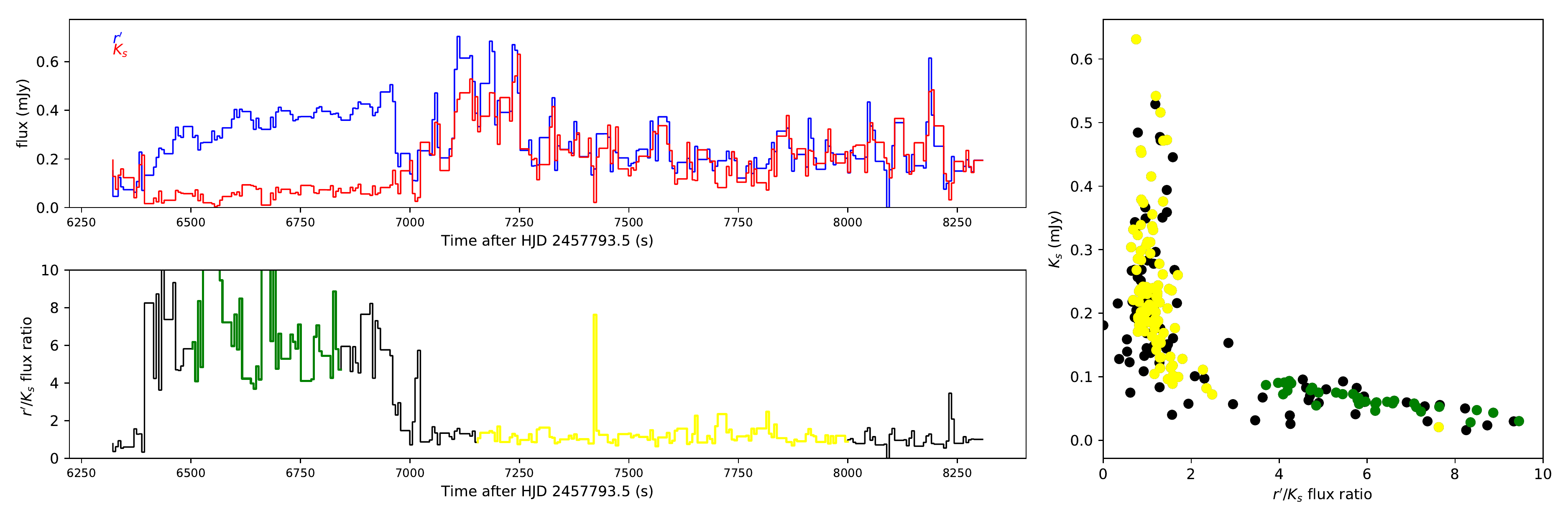}
\includegraphics[width=1.0\linewidth]{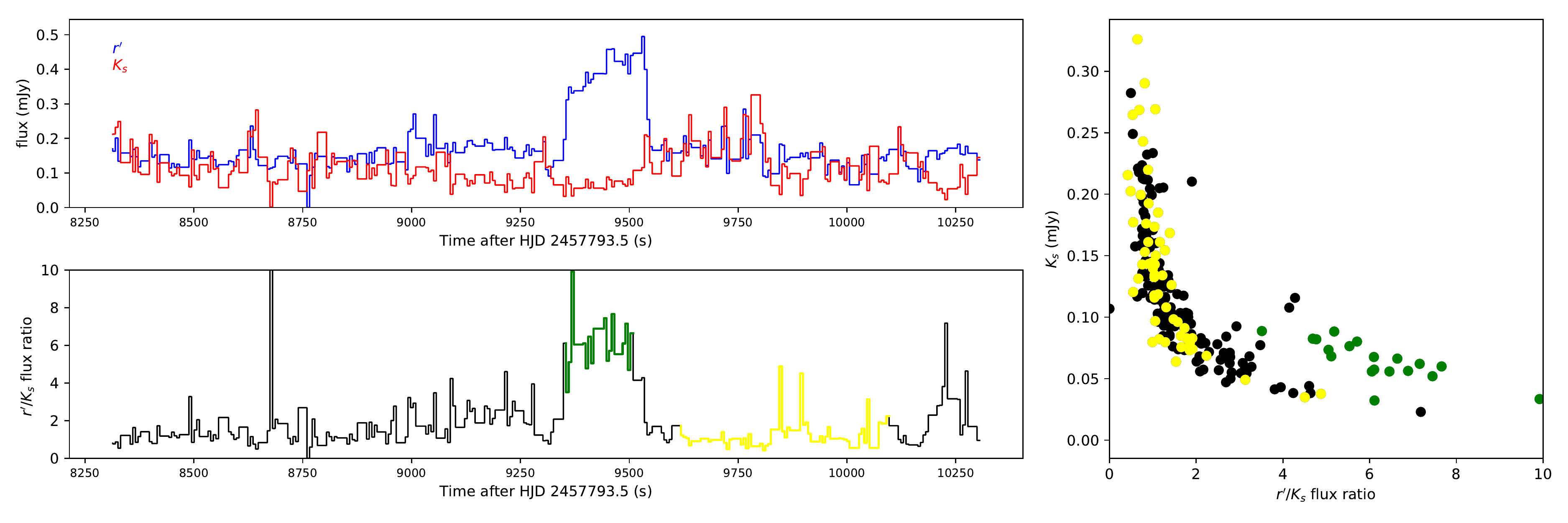}
\includegraphics[width=1.0\linewidth]{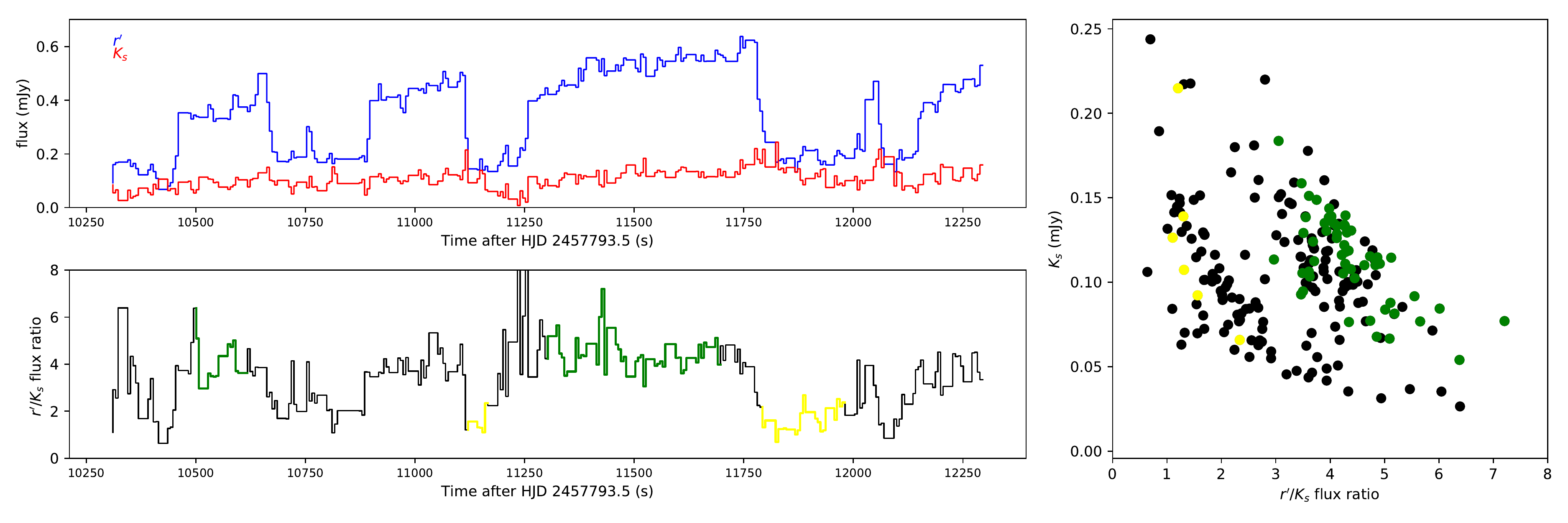}
\caption{
The light curves of  \target\ 
split into seven 2000\,s sections. Top left: The de-reddened and 
de-trended $r'$- and $K_s$-band light curves. Bottom left: The $r'$/$K_s$ 
flux ratio light curve. Right: The $r'$/$K_s$ versus $K_s$ flux ratio 
diagram. The yellow and green points mark examples of passive and 
active-state, respectively, which are shown in the $r'$/$K_s$ light curve plot (bottom 
right).
}
\label{fig:chunksA}
\end{figure*}

\section{LIGHT CURVES}
\label{LC:lcurves}

Figure\,\ref{fig:lcurve} shows the observed flux $r'$- and $K_s$-band light 
curves of \target\ and a comparison star of similar brightness, where we 
converted the magnitudes to flux density using the appropriate zero point 
for each band. The $r'$-band light curve clearly show the secondary star's 
heated ellipsoidal modulation with superimposed flaring activity. Flares 
and rectangular shaped dips are present similar to previous optical light 
curves \citep{Shahbaz15, Bogdanov15b}. We also show the light curves of the 
comparison stars. As one can see there is some low-level structure in the 
$K_s$-band comparison star light curve.

Fig.\,\ref{fig:detrend} shows the de-reddened, de-trended $r'$- and 
$K_s$-band flux light curves. We used a colour excess of E(B-V)\,=\,0.073 
\citep{Shahbaz15} and $A_{\rm V}/E(B-V)$\,=\,3.1 \citep{Cardelli89} to 
de-reddened the magnitudes and then converted the magnitudes to flux 
density. Finally we de-trended the light curves using a 3rd and 1st order 
polynomial for the $r'$- and $K_s$ band light curves, respectively. The 
passive-, active- and flare-states are clearly seen in the $r'$-band light 
curve rectangular, flat-bottomed dip features are apparent, which are 
similar to the mode-switching behaviour (passive- and active-state) that 
have been observed in the X-ray and optical light curves 
\citep{Tendulkar14,Bogdanov15b,Shahbaz15}. The histogram of the 
de-reddened, de-trended $r'$-band flux values (see Fig.\,\ref{fig:detrend}) 
clearly show a bi-modal distribution between the passive- and 
active-states. Such a distribution is not so apparent in the $K_s$-band 
light curve. This is because the rectangular dips in the $r'$-band light 
curve are deeper than the dips in the $K_s$-band light curve; the 
active minus passive (active-passive)  flux of the de-reddened light curves are 0.40 and 0.14\,mJy 
for the $r'$- and $K_s$ band light curves, respectively.

\begin{figure*}
\centering{
\includegraphics[width=1.0\linewidth]{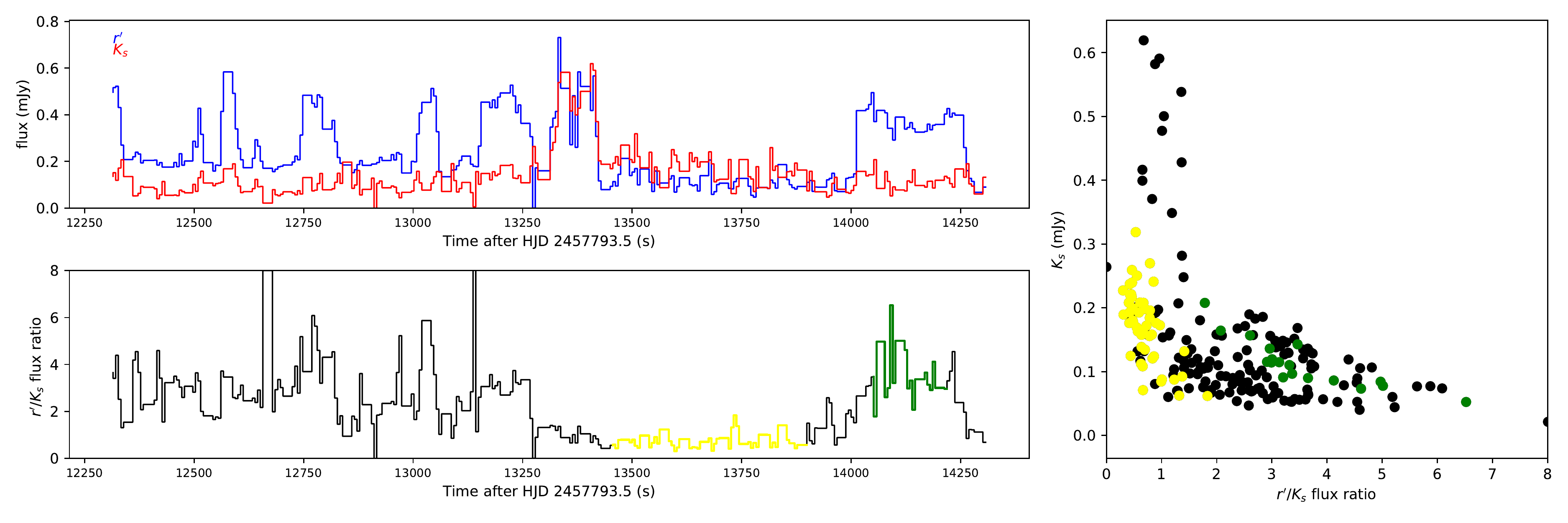}
\includegraphics[width=1.0\linewidth]{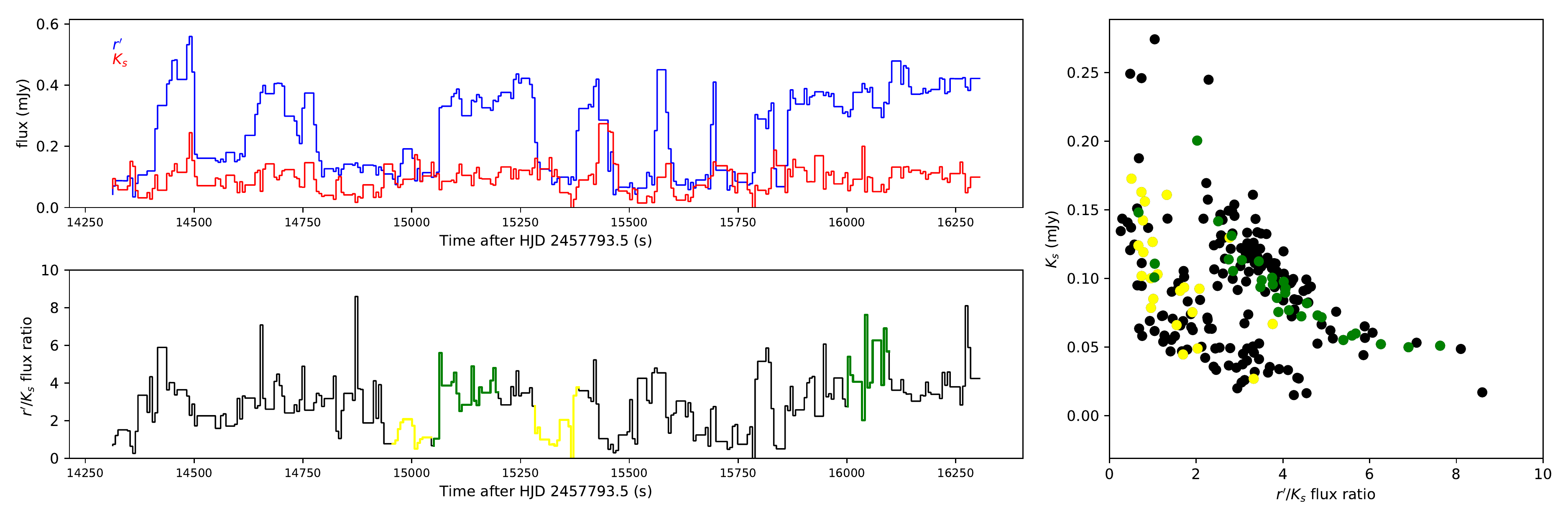}
\includegraphics[width=1.0\linewidth]{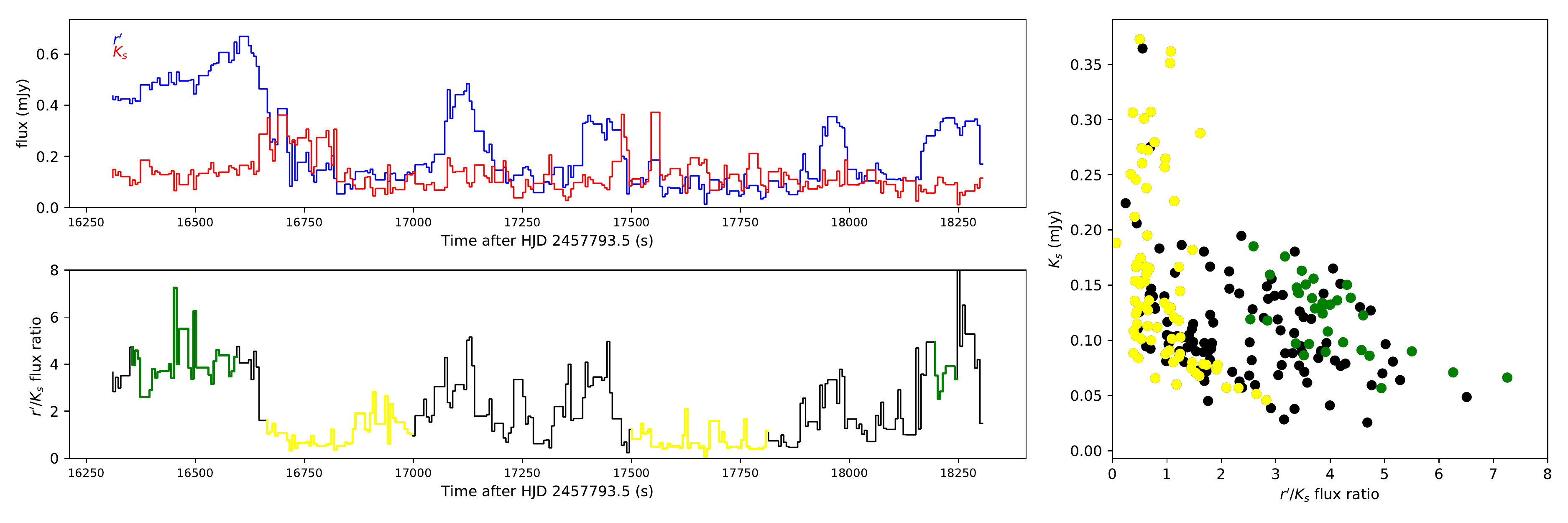}
\includegraphics[width=1.0\linewidth]{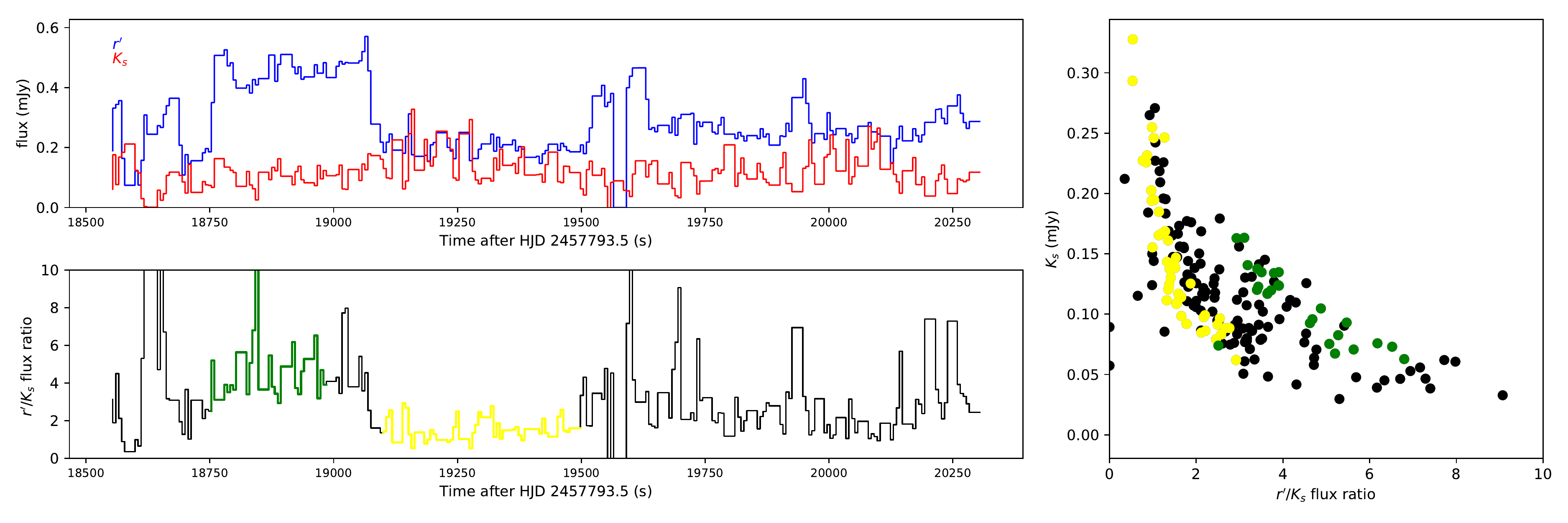}
}
\caption{
Same as Fig.\,\ref{fig:chunksA}.
}
\label{fig:chunksB}
\end{figure*}

\subsection{Light curve correlations}

In an attempt to understand the origin and emission mechanism of the 
optical and near-IR light curves, we calculate the ACF of the optical and 
near-IR light curves and the CCF between the observed optical and near-IR 
light curves. The analysis is a simple determination of the relative 
time-variability between the two wavelength dependent light curves and 
provides a clear, uncomplicated (average) comparison of the two signals 
with respect to time. The ACF represents a correlation of a light curve 
with itself, and gives a measure of the effective coherence time-scales in 
the data. The CCF represents the convolution of two light curve and 
reveals whether they have the same features and the time delay between 
them. To determine the ACF and CCF between two light curves (i.e. the 
cross-correlation coefficient versus time-delay between the light curves) 
we compute the discrete correlation function (DCF) using 
\textsc{pydc}\footnote{https://github.com/astronomerdamo/pydcf}, which 
works well with unevenly sampled data \citep{Edelson88,Robertson15}. 
Positive time-lags mean that the optical leads the near-IR, whereas 
negative time-lags means that the optical lags the near-IR. To 
calculate the ACF we use the whole light curve and the time-resolution of 
the data. To calculate the CCF between optical and near-IR light curves we 
only use data that are simultaneous (14000\,s). Given that time-resolution 
of the $r'$-band data is better than the $K_s$-band data, when calculating 
the CCF we rebin the light curves to the best time resolution of the 
$K_s$-band data. 

\subsubsection{ACFs}
\label{LC:corel}

In Fig.\,\ref{fig:ccfs_target} we show the ACF of entire $r'$- and 
$K_s$-band light curves of \target. The $r'$- and $K_s$-band ACFs have 
different widths; with a full width at half maximum (FWHM) of 130\,s and 
90\,s, respectively. Above $\sim$70\,s the $K_s$-band ACF is broader than 
the $r'$-band ACF, which argues for a reprocessing origin for the 
slowly-variable component, most likely in the outer regions of the 
accretion disc. 
However, very close to the core, the $r'$-band ACF is broader than the 
$K_s$-band ACF, which argues against a simple reprocessing origin (see 
Section\,\ref{DISC:lcurves}). The $K_s$-band comparison star light curve 
(see Fig.\,\ref{fig:lcurve}) shows some low-level correlated variability. 
Indeed, this is seen in its ACF (see Fig.\,\ref{fig:ccfs_cstar}), which 
has a non-zero width, with a full width at half maximum of 40\,s, similar 
to the time-scale of the dither pattern (see Section\,\ref{obs:ir}). In 
contrast, the ACF of the $r'$-band comparison star light curve does not 
show any peaks. This suggests that there are some low-level systematics in 
the data reduction of the near-IR data possibly due to under-corrected 
cosmetic defects in the CIRCE engineering-grade detector array. Hence, the 
true width of the ACF of \target\ is smaller.

In Fig.\,\ref{fig:ccfs_target} we show the optical and near-IR CCF using 
all the simultaneous optical and near-IR data (14000\,s). The negative lag 
implies the optical band is delayed with respect to the near-IR, whereas 
an anti-correlation implies that the variability between the near-IR and 
optical fluxes is out of phase. We observe a strong, broad negative 
anti-correlation on time-scales of $\sim$-300\,s, a broad positive 
correlation near-IR lag at $\sim$+300\,s and a strong, positive and narrow 
correlation at $\sim$+5\,s (see Fig.\,\ref{fig:ccfs_target}).

To see if there is a time-dependent evolution in the ACF and CCF
we compute the dynamic ACF and CCFs (see Fig.\,\ref{fig:dyn_acf})
calculated using a windows size of 1000\,s and a step size of 500\,s, i.e. 
overlapping sections where every second ACF/CCF is independent. A mean CCF 
is then derived for each time series pair, which effectively acts as a 
high-pass filter, and variations with time-scales longer than $\sim$500\,s 
do not appear. One can see, the $K_s$-band ACF is not always broader 
than the $r'$-band ACF, the width of the ACFs change with time.

\subsubsection{Light curves}
\label{LC:colcol}

In Figs.\,\ref{fig:chunksA} and \ref{fig:chunksB} we show the 
de-trended light curves and the $r'$/$K_s$ flux ratio and the $r'$/$K_s$ 
versus $K_s$-band flux ratio diagram. for each 2000\,s block of data. 
The yellow and green points mark the passive and active-states, respectively, which 
are also shown in the $r'$/$K_s$ light curve plot. During the transition 
from the active to passive-state, the $r'$/$K_s$ flux ratio decreases 
because there is a larger decrease in the $r'$-band flux compared to the 
$K_s$-band flux: there is a removal of a blue spectral component.

\subsection{POWER DENSITY SPECTRUM}
\label{PDS:pds}

\begin{figure}
\centering
\hspace{-5mm}
\includegraphics[width=1.05\linewidth]{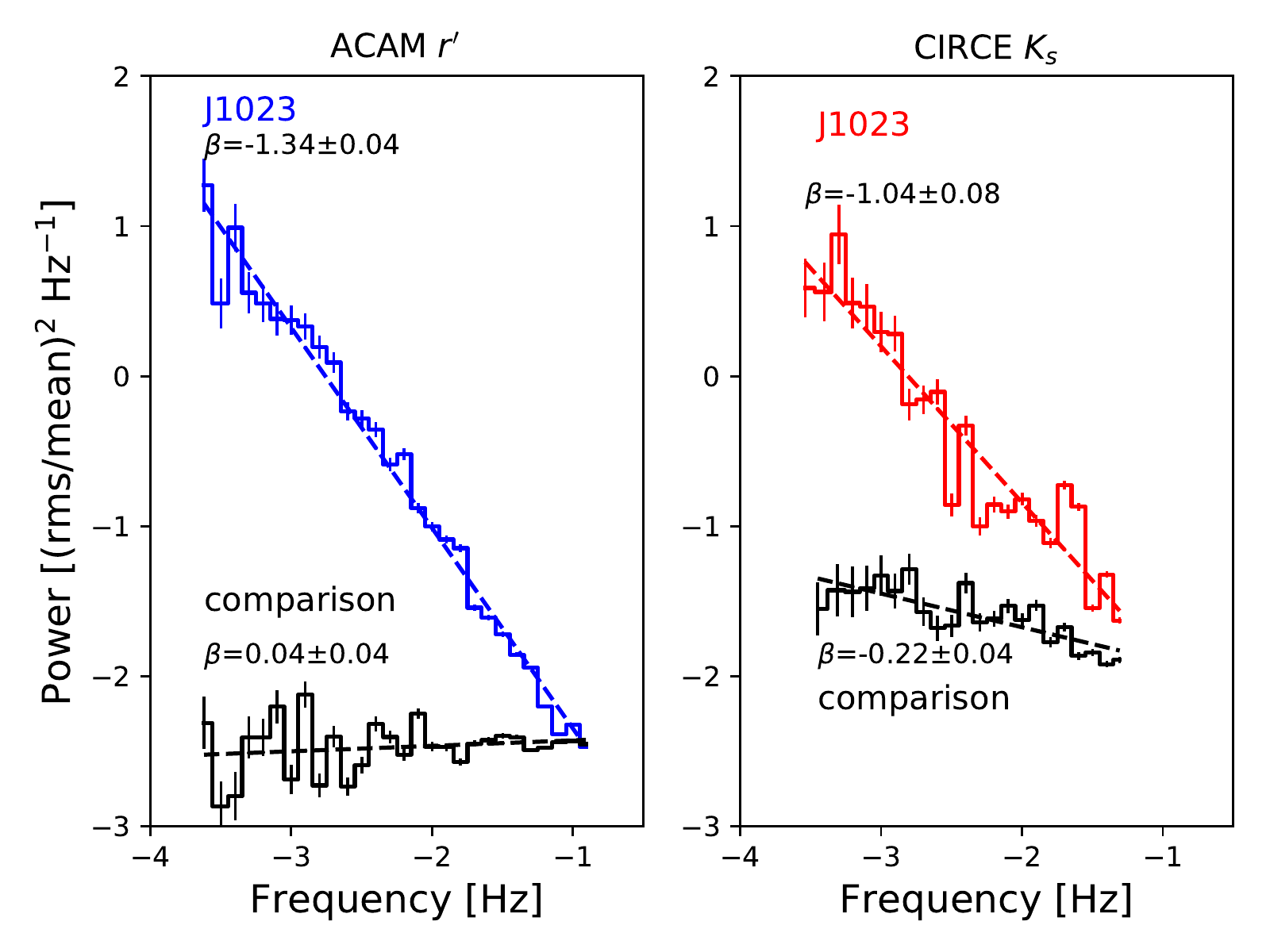}
\caption{
The PDS of the $r'$ (left) and 
$K_s$-band (right) light curves of \target\ and the comparison star. In 
each plot the dot-dashed lines show a power-law fit to the PDS. 
}
\label{fig:pds}
\end{figure}

In Fig.\,\ref{fig:pds} we show power density spectrum (PDS) of the 
de-reddened optical and near-IR light curves of \target\ (see 
Section\,\ref{LC:lcurves}). We use the Lomb-Scargle method to compute the 
periodograms \citep{Press92} with the constraints imposed by the Nyquist 
frequency, the typical duration of each observation and use the recipe 
given in \citet{Horne86} to calculate the number of independent 
frequencies. We then bin and fit the PDS in logarithmic space 
\citep{Papadakis93}, where white noise level was subtracted by fitting the 
highest frequencies with a white-noise (constant) plus red-noise 
(power-law; $P\propto\nu^\beta$) model (only for the \target\ light 
curves). We find that the PDS of the $r'$- and $K_s$-band light curve of 
\target\ is dominated by a red-noise component with a power-law index of 
$\beta=$-1.34$\pm$0.04 and -1.04$\pm$0.08, respectively, typical of 
aperiodic activity in X-ray binaries and X-ray transients in outburst and 
quiescence, which have $\beta$ in the range -1.0 to -2.0 \citep{Zurita03, 
Shahbaz03, Hynes03, Shahbaz04, Shahbaz05, Shahbaz10, Shahbaz13}. The 
power-law index of the $r'$- and $K_s$-band comparison star PDS 
$\beta$=0.04$\pm$0.04 and -0.22$\pm$0.04, respectively. The $r'$-band PDS 
is consistent with zero, however, for the $K_s$-band, the power-law index 
is significant, suggesting that there are systematic effects in the data 
(the PDS of the light curves using of a different local standard is also 
non-zero),  possibly due to under-corrected cosmetic defects in the 
CIRCE engineering-grade detector array.

\begin{figure}
\centering
\includegraphics[width=0.9\linewidth]{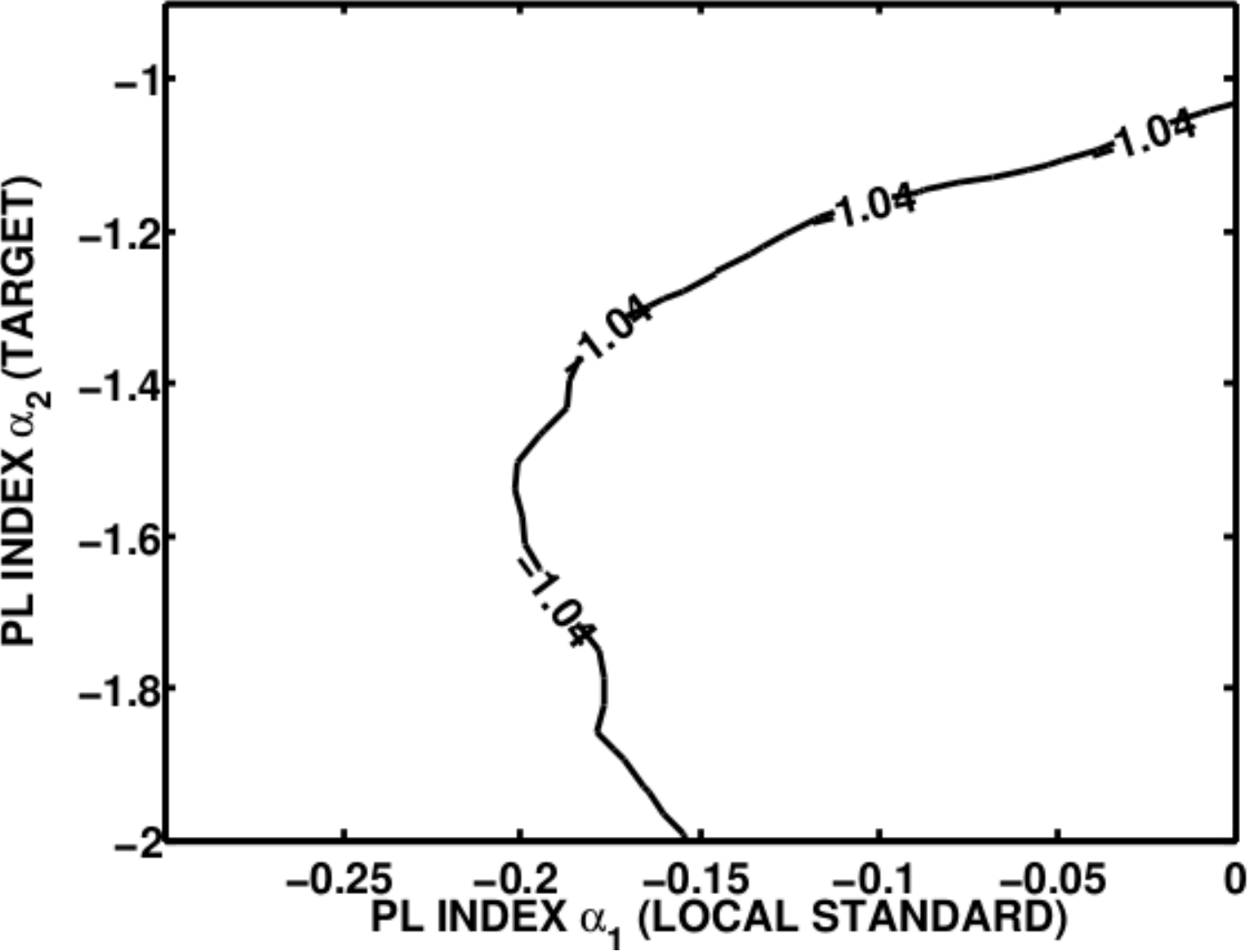}
\caption{
The effects of red-noise on the differential light curve of a 
target and local standard light curve. The light curves are calculated 
using using the method of \citet{Timmer95}, with different PDS power-law 
indices, $\beta_1$ and $\beta_2$ for the target and local standard light 
curve, respectively. The resulting differential light curve has a PDS 
power-index of $\beta$. The plot shows a contour where $\beta$=-1.04. The 
true PDS power-law index ($\beta_2$) of the target light curve is steeper 
than the power-law index of differential light curve $\beta$.
}
\label{fig:width_acf}
\end{figure}

When determining the light curve of \target\ we divided the counts of 
\target\ into the counts of the local standard which does not vary To 
investigative the effects of using a red-noise dominated local standard 
star (which has some low level variability) when computing the 
differential light curve of \target\ and comparison star, we perform a 
Monte Carlo simulation where we generate light curves with exactly the 
same sampling and integration times as the \target\ data. For each Monte 
Carlo, we simulate two light curve (LC$_1$ and LC$_2$) using the method of 
\citet{Timmer95}, with different PDS power-law indices, $\beta_1$ and 
$\beta_2$, respectively. We divide the two light curves (LC=LC$_2$/LC$_1$) 
and calculate the PDS. We perform 1000 simulations, bin and fit the PDS in 
logarithmic space \citep{Papadakis93}, and then determine the  PDS 
power-law index $\beta$. We assume that the local standard (LC$_1$) has 
low-level variability with $\beta_1$=0.0 to 0.50, whereas, the target of 
interest (LC$_2$) is dominated by red-noise, where $\beta_2$=0.5--1.5.
Our simulations show that $\beta$ lies between $\beta_1$ and $\beta_2$, 
i.e. the red-noise dominated local standard light curve (LC$_1$) makes 
power-law index of the differential light curve (LC) less steep. In 
Fig.\,\ref{fig:width_acf} shows the results of a simulation where final 
light curve LC has $\beta$=-1.04. As one can see the real PDS power-law 
index ($\beta_2$) of the target light curve LC$_2$ is steeper than the 
power-law index of differential light curve LC, $\beta_2<\beta$.
Although we cannot determine the real power-law index of the PDS of the 
$K_s$-band light curve because we do know the PDS power-law index of 
the local standard star used in the near-IR \target\ analysis. All we can 
say is we say is that the near-IR light curve of \target\ has a PDS of 
$<$-1.04$\pm$0.08. Given that the PDS of the comparison star is
-0.22$\pm$0.04, at most we expect a similar amount of variability in the 
local standard star. Hence, we find that $r'$-band PDS is steeper than the 
$K_s$-band PDS. Both PDS have similar power at low-frequencies, however, 
the $K_s$-band PDS has more variability at high frequencies.

\section{DISCUSSION}

\subsection{The spectral energy distribution of the rectangular dips}
\label{DISC:dips}

To date there are three (PSR\,J1023+0038, M28I and XSS\,J12270--4859) 
transitional pulsars systems that transition between accretion-powered 
LMXB and rotation-powered radio pulsar states \citep{Archibald09, 
Papitto13, Bassa14}. Of these systems \target\ and XSS\,J12270--4859 show 
rapid ($<$10\,s) transitions from a high X-ray luminosity active-state to 
a low luminosity passive-state. In \target\ similar optical transitions 
are observed but on a slightly longer time-scale ($\sim$20\,s.
Simultaneous {\it XMM--Newton} X-ray and $B$-band optical light curves show X-ray 
dips that do not usually have corresponding optical dips 
\citep{Bogdanov15b}. However, given that the $B$-band dips are much weaker 
compared to in the $r'$-band \citep{Shahbaz15}
combined with the large uncertainties in the $B$-band data, one cannot
cannot rule out if the dips are present and one cannot rule out that 
the X-ray and optical dips are direct counterparts of each 
other.

\citealt{Shahbaz15}) therefore suggest that the optical analogue of the X-ray 
model-switching may be due to reprocessing of the X-ray signal.

During the active-state X-ray pulsations are observed, which implies that 
matter is being channelled onto the neutron star's magnetic poles. 
However, during the passive-state there is no sign of X-ray pulsations 
\citep{Bogdanov14, Archibald15, deMartino13, Papitto15}. Indeed, optical 
pulsations have also been observed which are interpreted as being due to 
synchrotron emission by relativistic electrons and positrons in the pulsar 
magnetosphere \citep{Ambrosino17}. In the proposed model, the particles residing at 
a constant fraction of the light cylinder radius have a power-law energy 
distribution and emit incoherent synchrotron radiation \citep{Pacini83}. 
The synchrotron self-absorption produces a turnover in the 
spectral energy 
distribution (SED) Indeed, in the Crab there is evidence for synchrotron 
self-absorption which is observed as a turnover in the SED near the $H$-band \citep{O'Connor05}.

Simultaneous X-ray and radio observations have revealed key features 
\citep{Bogdanov17}. Firstly, an anti-correlation between the X-ray and 
radio flux in the active- and passive-states. During the X-ray active 
state and the radio luminosity is low and X-ray pulsations are observed, 
but when the source transitions into the passive-state, the pulsations 
disappear and the radio emission increases rapidly. Secondly, during the 
X-ray passive-state, the radio spectrum evolves from a self-absorbed to 
optically thin spectrum on time-scales of several minutes, reminiscent of 
a classical evolving synchrotron spectrum. \citet{Bogdanov17} conclude the 
radio flares are due to expanding synchrotron emitting plasma arising from 
the inner regions of the accretion disc.

\begin{figure}
\centering
\includegraphics[width=1.0\linewidth]{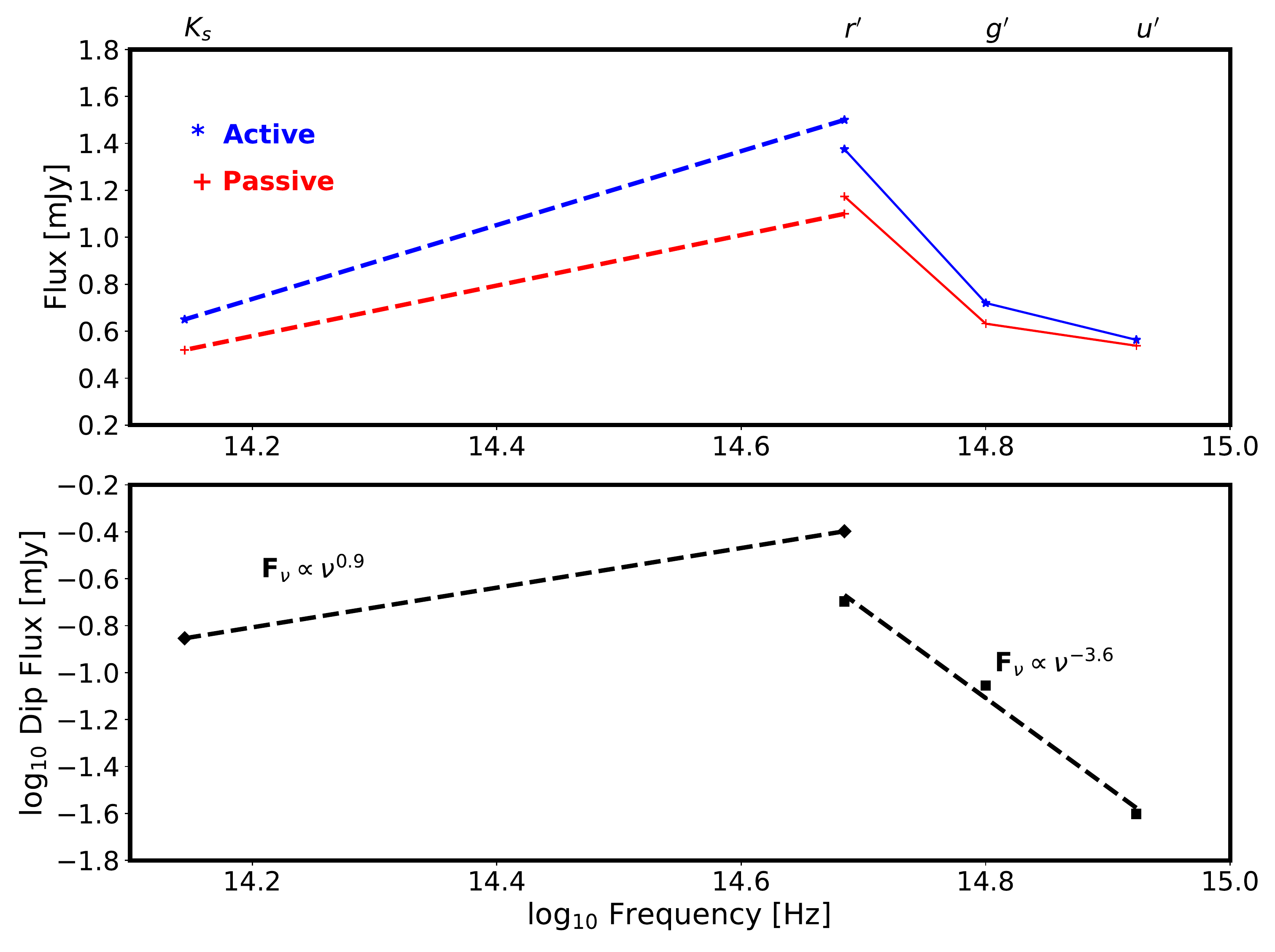}
\caption{
Top: the SED of the active- (blue stars) 
and passive- (red crosses) state. The $u'g'r'$-band data (solid line) were taken 
in 2015 and are presented in \citet{Shahbaz15} and the $r',K_s$ points (dashed 
line) were taken in 2017 presented in Section\,\ref{LC:lcurves}. Bottom: The 
active minus passive (active-passive) flux SED. The filled squares 
and filled diamonds show the data taken in 2015 and 2017, respectively. 
The dashed line shows a power-law fit of the form $F_\nu \propto 
\nu^\alpha$. A turnover in the SED lies somewhere 
between the $r'$- and $K_s$-band. The active and passive flux are 
de-reddened and contain flux from the secondary star, accretion disc and 
synchrotron emission of plasmoids and reprocessing 
(see Section\,\ref{DISC:flow}).
}
\label{fig:sed}
\end{figure}

In Fig.\,\ref{fig:sed} we show the SED of the 
$u'$-, $g'$- and $r'$-band dips observed in 2015 by \citet{Shahbaz15} and 
the $r'$- and $K_s$-band dips observed in 2017 (see 
Section\,\ref{LC:lcurves}). The SED shows a 
turnover which is what one expects if the synchrotron emission arises from 
an expanding synchrotron emitting plasma \citep{Bogdanov17} or synchrotron 
emission by relativistic electrons and positrons \citep{Ambrosino17}. The 
synchrotron emission idea is also supported by the high $K_s$-band PDS 
variability at high frequencies (inner disc regions; see 
Section\,\ref{PDS:pds}), which is what one expects for emission is 
non-thermal, likely from optically thin synchrotron emission.

An alternative explanation for the turnover in the SED 
of the dips is from a combination of clumpy accretion and 
irradiation. In Figs.\,\ref{fig:chunksA} and \ref{fig:chunksB} we show the 
$r'$/$K_s$ versus $K_s$-band flux ratio diagram for different sections of the 
data. As one can see, during the transition from the active to 
passive-state, the $r'$/$K_s$ flux ratio decreases because there is a 
larger decrease in the $r'$-band flux compared to the $K_s$-band flux, i.e there is 
a disappearance of a blue component. This is what one would expect if 
the reprocessing optical component is removed during the active--passive state transitions while 
the synchrotron component remains constant (see Section\,\ref{DISC:flow}).  
In  contrast \citet{Shahbaz15} observed the disappearance of a red spectral 
component and attributed it to the removal of an irradiated disc. 
The difference between the $r'$-band dip observed in 2015 and 
2017 (see Fig.\,\ref{fig:sed}) is due to the different amounts of clumpy 
material at the time of the observations. The combination of the removal 
of a red and blue component with different strengths gives rise to a 
turnover in the SED which lies somewhere between the 
$r'$- and $K_s$-band.

\subsection{Comparison with X-ray binaries}
\label{DISC:xrb}

Simultaneous optical and X-ray observations of X-ray binaries have 
revealed multi-component optical variability on sub-second time-scales, 
arising from components such as a hot inner flow, accretion disc 
reprocessing and a jet. These appears to be a common feature in many X-ray 
binaries during the hard state and resuls in as narrow positive 
peak is the observed in the CCF, superimposed on an anti-correlation 
\citep{Durant11, Durant08, Gandhi08, Malzac03, Kanbach01}. Intense and 
rapid sub-second flaring is also seen and in a few systems it has been 
attributed to synchrotron emission from a jet \citep{Gandhi10, Malzac04, 
Hynes03}. On time-scales of minutes Swift\,J1753.5−0127 has shown strong 
optical variations \citep{Durant09}. \citet{Veledina15a,Veledina17} have 
shown that the X-ray and optical variability in Swift\,J1753-0127 
\citep{Durant08} is well reproduced by a simple model assuming that the 
optical variability is a mixture of X-ray reprocessing plus a hot flow 
synchrotron component that is exactly anti-correlated with respect to the X-rays. 
The hot flow model has also been used to explain the anti-correlated 
optical/X-ray behaviour in BW\,Cir \citep{Pahari17}. Near-IR variations on 
time-scales of minutes have been observed in GRS\,1915+105 and have been 
associated with synchrotron-emitting plasma \citep{Fender97, 
Eikenberry98}.

The shape of the optical/near-IR CCF of \target, with the anti-correlation 
and a sharp rise at zero lag, resembles the optical/X-ray CCFs found in 
X-ray binaries, indicating an interplay of, at least, two separate 
components. This scenario was originally proposed for black hole X-ray 
binaries \citep{Veledina11}, where the correlated component comes from the  
reprocessing of the incident X-ray flux, and the anti-correlated component 
originates from the hot accretion flow emitting in the optical via 
synchrotron. The time-scale of the optical and near-IR variability and 
correlations we observe in \target\ are, however, much longer than what is 
observed in other X-ray binaries.

A hot accretion flow can, in principle, can exist close to the pulsar 
magnetosphere and can be occasionally expelled away by the rotating 
magnetic field \citep{Parfrey17}. Such synchrotron-emitting outflowing 
plasmoids arising from the inner regions of the accretion disc could then 
be responsible for the increased radio emission during the X-ray passive 
states \citep{Bogdanov17}. Given the low luminosity, and hence, low 
accretion rate and electron number density in \target\, the hot flow 
synchrotron emission is likely to be optically thin down to the near-IR. 
The entire spectrum is thus similar to those of the outer parts of the hot 
flow for higher accretion rates \citep{Veledina13, Poutanen14b}. A 
combination of the red synchrotron spectrum with the blue spectrum of 
reprocessed radiation (Veledina et al., 2018, in preparation) can produce 
the observed power-law $F_{\nu} \propto \nu^{0.9}$ (see Fig.\,\ref{fig:sed}).

The shape of the optical/near-IR CCF of \target\ somewhat resembles the 
optical/X-ray CCF observed in the black hole X-ray binary 
Swift\,J1753.5--0127 \citep{Durant08}. However, the characteristic 
time-scales are different by an order of magnitude, and the broad positive 
peak at positive lags is missing in the CCF of Swift\,J1753.5--0127. The 
difference can be explained by the presence of a solid surface and neutron 
star magnetosphere in \target. Because the accretion disc is truncated at this 
radius, the characteristic viscous time-scales are somewhat longer than in 
black hole binaries, giving broader features in the CCF. The absence of 
the broad peak at positive lags can be due to disappearance of the 
plasmoids under the event horizon, rather than their ejection, as is the 
case of \target.

\begin{figure*}
\centering
\includegraphics[width=1.0\linewidth]{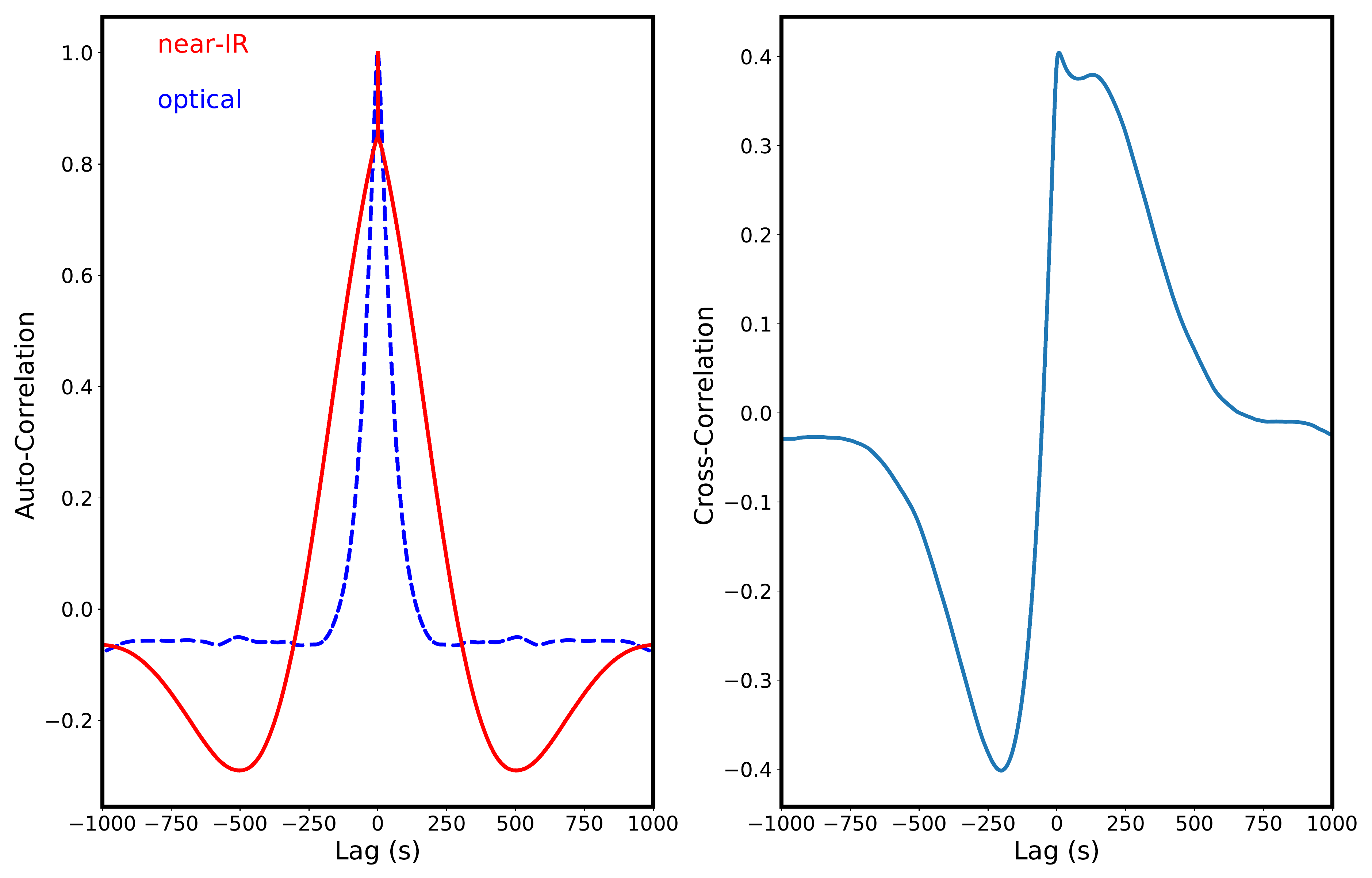}
\caption{
Model predictions for the optical and near-IR ACFs and CCF. 
respectively Left: the optical (blue dashed line) and near-IR (red solid 
line) ACFs. Right: the optical and near-IR CCF. The model considers a hot 
accretion flow of \citet{Veledina11} with the inner radius close to the 
magnetospheric radius of the neutron star. The near-IR/optical CCF can be 
explained by two components, synchrotron emission of plasmoids and 
reprocessing. The dip at negative lags corresponds to suppression of the 
near-IR synchrotron emission of the plasmoids in hot accretion flow. The 
narrow peak at $\sim$5~s corresponds to the delayed reprocessed component, 
enhanced by the increased X-ray emission. The broad positive correlation 
at positive lags is driven by the increased synchrotron emission of the 
outflowing plasmoids. 
}
\label{fig:model}
\end{figure*}

\subsection{Light curve correlations}
\label{DISC:lcurves}

Reprocessing of high energy photons to lower energies in an extended 
accretion disc is thought to dominate the observed optical and near-IR 
fluxes. The expected cross-correlation signature is delayed because of 
the light travel time and smeared response of the optical and near-IR 
light curves with respect to the X-ray light curve. The optical and near-IR 
light curve is the convolution of the X-ray light curve with a broad 
transfer function representing the response of the extended accretion 
disc. The inner regions of the accretion disc reprocess high energy X-rays 
into the soft X-rays/UV bands. Therefore we expect the near-IR ACF to be 
broader than the optical CCF and the CCF to show a single strong, positive 
peak at positive lags (near-IR arriving after optical).

At first sight the ACFs of the entire $r'$- and $K_s$-band light curves 
have different width cores and wings (see Section\,\ref{LC:corel}). Near 
the the $K_s$-band ACF is broader than the $r'$-band ACF which argues for 
a reprocessing origin for the slowly-variable component, most likely in 
the outer regions of the accretion disc. However, very close to the core, 
the $r'$-band ACF is broader than the $K_s$-band ACF, which argues against 
a simple reprocessing origin. A detailed look at the ACF for different 
segments in time reveals that on average the $r'$-band ACF is a factor of 
$\sim$3--4 times broader than the $K_s$-band ACF (see 
Fig\,\ref{fig:dyn_acf}), which is not what one would expect from a 
reprocessing model. A broader $K_s$-band ACF is expected if the $K_s$-band 
PDS is steeper than the $r'$-band PDS \citep{Peterson01} which implies 
that there is more aperiodic variability from the accretion disc at low 
frequencies (outer disc regions) in the $K_s$-band light curve compared to 
the $r'$-band. However, we find that the $r'$-band PDS is steeper than the 
$K_s$-band PDS (even allowing for the systematic effects, see 
Section\,\ref{LC:corel}) and that the $K_s$-band PDS has more variability 
at high frequencies from the inner regions of the accretion disc/pulsar 
magnetosphere (see Section\,\ref{PDS:pds}). This is what one expects for 
emission from a non-thermal component such as optically thin synchrotron 
emission.

The CCF of the simultaneous optical and near-IR data reveals a strong, 
broad negative anti-correlation and negative lags and a broad positive 
correlation at positive lags, with a strong, positive and narrow 
correlation superimposed (see Fig.\,\ref{fig:ccfs_target}). The broad 
positive correlation implies that the optical leads the near-IR flux. 
However, the broad negative delay implies the optical band is delayed with 
respect to the near-IR and the anti-correlation implies that the 
variability between the near-IR and optical fluxes is out of phase.

The narrow correlation implies that optical leads the near-IR flux, 
consistent with a reprocessing origin. The $\sim$+5\,s correlation is 
similar to the time-scale we would expect for the light travel time 
between the binary components. The light travel times of photons arise 
from the differences between the time of flight of photons that are 
observed directly and those that are re-processed and re-emitted before 
travelling back to the observer. For \target\ with an orbital period of 
4.75\,h, the corresponding binary separation of $\sim$1.8\,$R_{\odot}$ 
(assuming a total binary mass of 2\,$M_{\odot}$)  implies a light travel 
time of $\sim$4 light-seconds. However, given that the delay is not phase 
resolved, it is not always present in the CCFs (see 
Figs.\,\ref{fig:chunksA} and \ref{fig:chunksB}), suggesting that it is not 
due to the reprocessing on the secondary star, but is due to 
reprocessing in the accretion disc.

\subsection{The hot accretion flow model}
\label{DISC:flow}

The observed broad anti-correlation at negative lags implies that 
the optical flux is delayed with respect to the near-IR flux and the 
anti-correlation implies that the variability between the near-IR and 
optical fluxes is out of phase. These features are not due to reprocessing 
because a reprocessing CCF would rise quickly near zero and fall less 
quickly towards positive lags \citep{O'Brien02}. The broad components of 
the CCF and the sudden change near zero lag are strongly indicative of 
at least two separate processes superposed in the light curves, such as a 
hot flow for the anti-correlation and a reprocessing/jet component for the 
positively correlated optical/near-IR components. A combination of 
reprocessing and the interaction of a hot flow with the pulsar 
magnetosphere can produce the observed optical/near-IR correlations.

Given that the reprocessed emission is expected to have a blue spectrum, 
we expect it to be more prominent in the $r'$-band compared to in the 
$K_{\rm s}$-band. We propose that the $r'$-band emission is totally 
dominated by the reprocessed component, while the $K_{\rm s}$-band 
contains both synchrotron and reprocessed emission. 
In the optical the reprocessing component dominates in the 
active--state transiton and decreases during the transition. In the 
near-IR, there are two components, a reprocessing component which also 
decreases (but not as much as in the optical) during the transition, however, 
the synchrotron does not, which results in the removal of the blue spectral 
component during the dips (see Figs.\,\ref{fig:chunksA} and \ref{fig:chunksB}). 
If the accretion proceeds through the occasional penetration 
of the clumps/blobs of plasma through the magnetic barrier, the following 
scenario can lead to the observed near-IR/optical CCF. Consider the hot 
accretion flow with the inner radius close to the magnetospheric radius 
and emitting through the synchrotron self-Compton mechanism (similar to 
the hot flows in black hole X-ray binaries \citep{Poutanen14a}. If there 
is a local increase in the mass transfer rate, then the accretion flow 
becomes inhomogeneous and the resulting spectrum depends on the local 
number density of electrons. For a higher electron number density the 
(near-IR) synchrotron emission is suppressed due to the increased 
self-absorption. The inhomogeneities in the hot accretion flow containing 
a higher electron number density have a higher chance to penetrate through 
the magnetic barrier, hence once there is a sufficiently dense plasmoid, 
with the near-IR synchrotron emission suppressed, the flow accretes/drifts 
towards the compact object causing an increase in the X-ray emission. The 
X-ray emission triggers the reprocessing in the disc, both in the optical 
and near-IR (likely without a significant delay between them). The 
plasmoids are then expelled by the rotating magnetic field, forming an 
outflow and giving rise to the increase of the synchrotron near-IR 
emission as the plasmoids become optically thin. At the same time, the 
X-ray luminosity drops and the source switches to the passive state. 
Indeed, these ideas are further supported by the fact that transition from 
the X-ray active to X-ray passive state are always accompanied by a radio 
brightening, which has been interpreted as being due to rapid ejections 
synchrotron emitting plasma by the pulsar \citet{Bogdanov17}.

The near-IR/optical CCF can be explained by two components, synchrotron 
emission from plasmoids and reprocessing. The dip at negative lags 
corresponds to the suppression of the near-IR synchrotron emission of the 
plasmoids in hot accretion flow, which occurs just before the X-ray switches 
to the active-state. The narrow peak at $\sim$5\,s corresponds to the 
delayed reprocessed component, enhanced by the increased X-ray emission 
\citep{O'Brien02}. The broad positive correlation at positive lags is 
driven by the increased synchrotron emission of the outflowing plasmoids, 
which become optically thin as they expand. The width of the 
anti-correlation and broad positive correlation corresponds to the 
characteristic time-scale for the penetration of the (entire) inhomogeneity 
into the magnetosphere, which is roughly comparable to the viscous time-scale at 
the magnetospheric radius. In the $r'$-band light curves, this corresponds 
to the average duration of the active-state.

In Fig.\,\ref{fig:model} we show model predictions for the optical and 
near-IR ACFs and CCFs expected from the hot accretion flow model. Our aim 
is to show that the model can indeed reproduce the main features of the 
CCFs. We leave the detailed modelling to a future paper, as it involves a 
number of unknown parameters for the disc and synchrotron components. To model the 
ACFs and CCFs, we simulate the light curve as mass accretion rate 
fluctuations from a prescribed PDS shape using the \citet{Timmer95} 
algorithm.  We assume that the X-ray light curve is proportional to the 
mass accretion rate light curve. We assume a zero-peak Lorentzian profile 
for the PDS accretion rate and equal contributions of the disc and 
synchrotron components. The disc to synchrotron ratio determines the 
relative importance of the dip and peak in the CCF \citep{Veledina17}. 
The $r'$-band light-curve is obtained by convolving the X-ray light curve 
with the accretion disc transfer function. The synchrotron emission of 
the plasmoids is simulated through the mean-subtracted accretion rate light 
curve. For the radiation of the hot flow we take this light curve with a 
minus sign (similar to eq. 2 of \citealt{Veledina11}) and shift it in time so 
that it leads the X-rays. The synchrotron emission of expanding plasmoids 
is modelled as the accretion rate light curve, shifted in time towards 
positive lags, i.e. it comes after the X-rays. The $K_{\rm s}$ band 
contains the synchrotron components of the plasmoids and the reprocessed 
component. The main parameters effecting the shape of the CCF are the assumed 
PDS shape of the mass accretion rate fluctuations and relative contribution of 
various synchrotron/disc components. The model predictions for the optical 
and near-IR ACFs and CCFs expected are 
shown in Fig.\,\ref{fig:model}. As one can see there is a remarkable 
agreement in the general features predicted by the model with what is 
observed, a broad negative anti-correlation at negative lags, a broad 
positive correlation at positive lags, with a strong, positive narrow 
correlation superimposed.

If the active-state luminosity $L_X \sim 3\times10^{33}$\erg\ 
\citep{Archibald15, Bogdanov15b} is due to accretion, i.e. 
$L\,=\,G\,M\,\dot{M}/R$, where the neutron star mass is $M=1.71M{\odot}$ 
\citep{Deller12}, radius $R=12$~km \citep{Nattila16} and $G$ is the 
gravitational constant, then the accretion rate is 
$\dot{M}\,=\,1.5\times10^{13}$$\rm \,g\,s^{-1}$. With this mass inflow 
rate at the magnetospheric boundary, the active-state magnetospheric 
radius (the radius when the pressure of the neutron star magnetosphere is  able
to halt the inflowing matter) is $R_{\rm m}\sim150$\,km, assuming a  dipole
magnetic field with strength $B\sim10^8$\,G \citep{Archibald09,Deller12}. In
the  passive-state, the mass inflow rate is likely smaller, leading to a
higher  magnetospheric radius.  The corotation radius (at which the matter in 
Keplerian orbit co-rotates with the neutron star) is $R_{\rm co}$=24\,km  and
the light cylinder radius (at which field lines anchored to the  neutron star
rotate at the speed of light) is $R_{\rm lc}$ = 80\,km.   Hence for such low
X-ray luminosities, the magnetospheric radius exceeds  the light cylinder radius, 
and by far exceeds the corotation radius. This implies that the hot flow has to
be truncated outside of the light cylinder, $\gtrsim$100\,km. The viscous
time-scale at this radius is of  the order of a few 100\,s, which is roughly
consistent with the width of the  negative dip and broad peak in the CCF.  The
magnetic field inside the  flow is likely self-generated by the flow and can be
much lower than   at the neutron star surface.  We estimate the magnetic field
inside the  flow using condition that the turnover frequency is close to
$K_s$-band  and the requirement for the observed synchrotron luminosity is equal
to  the synchrotron luminosity at the turnover frequency. Using eq.~(6) of 
\citet{Veledina13} for the $K_s$-band flux at a radius of $\sim $100\,km  and a
distance of 1.368\,kpc \citep{Deller12}, we obtain a rough estimate  for the
magnetic field strength inside the hot flow of $B\sim10^4$\,G, 
which is comparable to the magnetic field of the pulsar at this radius.
We leave the
detailed spectral and timing modelling of the observed  characteristics to a
future paper.

\section*{CONCLUSION}

Below we list the main results of this paper.

\begin{enumerate}
\item 
Our simultaneous $r'$- and $K_s$-band light curves  show
rectangular, flat-bottomed dips, similar to the X-ray mode-switching 
(active--passive state transitons) behaviour observed previously. During the 
active--passive state transition, the $r'$/$K_{\rm s}$ flux ratio  decreases due to
the larger decrease in the $r'$-band flux compared to  the $K_{\rm s}$-band flux.  The SED
of the dips (active minus passive flux) shows a  turnover between the $r'$- and
$K_{\rm s}$-band.

\item
The CCF of the optical and near-IR data reveals a strong, broad negative 
anti-correlation and negative lags, a broad positive correlation at 
positive lags, with a strong, positive narrow correlation superimposed. 
The shape of the CCF resembles the optical/X-ray CCF observed in the black 
hole X-ray binary Swift\,J1753.5--0127, but the time-scales are different. 
The difference can be explained by the presence of a solid surface and 
the pulsar's magnetosphere in \target.

\end{enumerate}

\noindent 
The active--passive transiton and features in the CCF can be 
explained by reprocessing and a hot accretion flow close to the 
neutron star's magnetospheric radius. Since the accretion disc is 
truncated at distance comparable to the magnetospheric radius, the 
characteristic viscous time-scales are longer than in black hole 
binaries, giving broader features in the CCF. The optical emission is 
dominated by the reprocessed component, whereas the near-IR emission 
contains a component due to the emission from plasmoids in the hot 
accretion flow and a reprocessed component. 

The rapid transitions between the active and passive states are 
due to the penetration of the hot accretion flow material onto the 
neutron star and the expelling of the blobs of synchrotron emitting 
plasma fed by the hot inner flow from the magnetosphere. The 
plasmoids are expelled by the rotating magnetic field of the neutron 
star, giving rise to the increase of the synchrotron near-IR emission 
as they expand and become optically thin. During the active--passive 
transition the optical reprocessing component decreases which results 
in the removal of a blue spectral component.
The combination of the red synchrotron spectrum with the blue 
reprocessed spectrum also produces the observed SED. The dip at 
negative lags corresponds to the suppression of the near-IR 
synchrotron component in the hot flow, whereas the broad positive 
correlation at positive lags is driven by the increased synchrotron 
emission of the outflowing plasmoids. The narrow peak in the CCF is 
due to the delayed reprocessed component, enhanced by the increased 
X-ray emission. The accretion of clumpy material through the magnetic 
barrier of the neutron star produces the observed near-IR/optical CCF 
and variability.

\section*{ACKNOWLEDGEMENTS}

TS acknowledges support from the Spanish Ministry of Economy and 
Competitiveness (MINECO) under the grant AYA2013-42627. AV acknowledges 
support from the Academy of Finland grant 309308. P.G. thanks STFC 
(ST/J003697/2) for support. The ACAM and CIRCE data are based on 
observations made with William Herschel Telescope and the Gran Telescopio 
Canarias, respectively, installed in the Spanish Observatorio del Roque de 
los Muchachos of the Instituto de Astrof\'\i{}sica de Canarias, in the island 
of La Palma. Development of CIRCE was supported by the University of 
Florida and the National Science Foundation (grant AST-0352664), in 
collaboration with IUCAA.

\medskip

\noindent
{\it Facilities:} WHT (ACAM), GTC (CIRCE)


\end{document}